\documentclass[aps,prb,preprint,letterpaper,amsfonts,amssymb,floatfix,amsmath,showpacs,showkeys,unsortedaddress]{revtex4}
\usepackage{natbib}
\usepackage{graphicx}
\newcommand{\LCMO}{$\text{La}_{1-x}\text{Ca}_{x}\text{MnO}_{3}$~}
\newcommand{\NPMO}{$\text{Nd}_{1-x}\text{Pb}_{x}\text{MnO}_{3}$~}
\newcommand{\NPMOsam}{$\text{Nd}_{0.7}\text{Pb}_{0.3}\text{MnO}_{3}$~}

\newcommand{\LCMOoneeight}{$\text{La}_{0.82}\text{Ca}_{0.18}\text{MnO}_{3}$~}
\newcommand{\LAMO}{$\text{L}_{1-x}\text{A}_{x}\text{MnO}_{3}$~}
\newcommand{\tc}{$\text{T}_{C}$~}
\newcommand{\tp}{$\text{T}_{p}$~}
\newcommand{\tfmi}{$\text{T}_\text{FMI}$~}

\newcommand{\ilow}{$\text{I}_\text{low}$~}
\newcommand{\ihigh}{$\text{I}_\text{high}$~}
\newcommand{\jlow}{$\text{j}_\text{low}$~}
\newcommand{\jhigh}{$\text{j}_\text{high}$~}
\newcommand{\jth}{$\text{j}_\text{th}$~}
\newcommand{\eg}{$\text{e}_{g}$~}
\newcommand{\mub}{$\mu_{B}$~}

\begin{document}
\title{Colossal electroresistance in ferromagnetic insulating state of single crystal \NPMOsam}
\date{\today}
\author{Himanshu Jain}
\email[Electronic mail: ]{himanshu@physics.iisc.ernet.in}
\affiliation{Department of Physics, Indian Institute of Science, Bangalore 560 012, India}
\author{A. K. Raychaudhuri}
\email[Electronic mail: ]{arup@bose.res.in}
\altaffiliation{On lien from: Department of Physics, Indian Institute of Science, Bangalore 560 012, India.}
\affiliation{S. N. Bose National Centre for Basic Sciences, Salt Lake, Kolkata 700 098, India}
\author{Nilotpal Ghosh}
\author{H. L. Bhat}
\affiliation{Department of Physics, Indian Institute of Science, Bangalore 560 012, India}

\begin{abstract}
Colossal electroresistance (CER) has been observed in the ferromagnetic insulating (FMI) state of a manganite. Notably, the CER in the FMI state occurs in the absence of magnetoresistance (MR). Measurements of electroresistance (ER) and current induced resistivity switching have been performed in the ferromagnetic insulating state of a single crystal manganite of composition \NPMOsam (NPMO30). The sample has a paramagnetic to ferromagnetic (Curie) transition temperature, \tc $= 150$ K and the ferromagnetic insulating state is realized for temperatures, T $\lesssim 130$ K. The colossal electroresistance, arising from a strongly nonlinear dependence of resistivity ($\rho$) on current density (j), attains a large value ($\approx 100\%$) in the ferromagnetic insulating state. The severity of this nonlinear behavior of resistivity at high current densities is progressively enhanced with decreasing temperature, resulting ultimately, in a regime of negative differential resistivity (NDR, d$\rho$/dj $<$ 0) for temperatures $\lesssim 25$ K. Concomitant with the build-up of the ER however, is a collapse of the MR to a small value ($< 20\%$) even in magnetic field, H $= 7$ T. This demonstrates that the mechanisms that give rise to ER and MR are effectively decoupled in the ferromagnetic insulating phase of manganites. We establish that, the behavior of ferromagnetic insulating phase is distinct from the ferromagnetic metallic (FMM) phase as well as the charge ordered insulating (COI) phase, which are the two commonly realized ground state phases of manganites.
\end{abstract}


\maketitle

\section{Introduction}
Electronic transport in hole doped rare earth manganites \LAMO (L $\equiv$ Nd, La, Pr; A $\equiv$ Pb, Ca, Sr, Ba; $\textit{x}\lesssim0.3$) is an issue of considerable interest since the observation of colossal magnetoresistance (CMR) in this class of materials~\cite{RaoBOOK1998,TokuraBOOK2000TokuraRPP2006}. These materials also exhibit colossal electroresistance (CER), wherein an applied  electric field (as in a field effect (FE) configuration~\cite{ZhaoAPL2004ZhangAPL2005,WuPRL2001,OdagawaPRB2004}) or an applied current (as in a typical 4-probe configuration~\cite{AsamitsuNATURE1997,DebnathPRB2003,ZhaoAPL2005}) can change the resistivity ($\rho$) of the sample significantly. Most investigations of CER, report a decrease of sample resistivity with increasing sample current. It would be worthwhile to have a clear understanding whether the CER and CMR effects have a common physical origin. We have recently reported that, in at least one system, namely  single crystals of composition \LCMOoneeight, the CER and CMR phenomena get completely decoupled when the system is in its ferromagnetic insulating (FMI) state~\cite{JainAPL2006}. (Note: By decoupling we mean that a CER is present in the absence of a CMR effect.) In this paper, we report further strong evidence in favor of different origins of the two phenomena in another and very different manganite system. The system reported in this paper is a single crystal belonging to the family \NPMO.

\begin{figure}[t]
\includegraphics{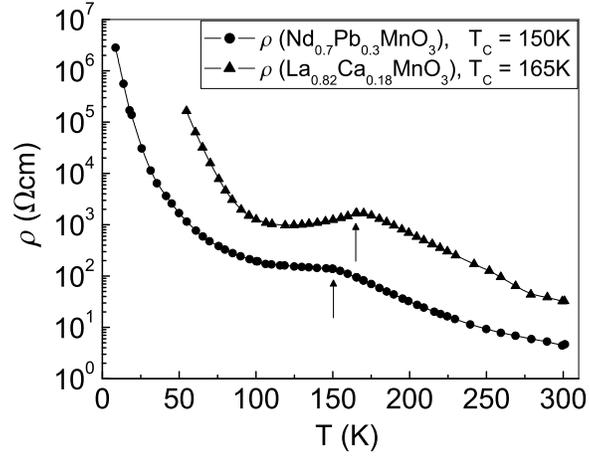}
\caption{\label{Fig1NPMO30LCMO18RhovsTComparison}Comparison of resistivities ($\rho$) of \NPMOsam and \LCMOoneeight as a function of temperature (T). The paramagnetic to ferromagnetic (Curie) transition temperatures (\tc), indicated by arrows, are $150$ K and $165$ K respectively.} 
\end{figure}

The phase diagram of the \NPMO system has been studied before and shows that, unlike the \LCMO system which shows FMI state for $x\lesssim 0.18$, the \NPMO system shows FMI behavior even upto a carrier concentration of $x\approx0.3$~\cite{GhoshJAP2004}. A likely cause of this behavior is the presence of a large disorder due to the very dissimilar ionic sizes of the two ions, Nd and Pb. This disorder can stabilize the insulating state to such high carrier concentration ($x\approx 0.3$). The existence of FMI behavior at such a large value of carrier concentration ($x\approx 0.3$) makes \NPMO a very different system from the much studied \LCMO system; this is apparent from the much reduced resistivity, in the FMI phase, of the composition \NPMOsam (NPMO30) as compared against that of the composition \LCMOoneeight (LCMO18). For instance, the room temperature resistivity of \LCMOoneeight, $\rho_\textrm{LCMO18}$(T $= 300$ K) $= 32$ $\Omega$cm which, deep in the FMI phase (T $= 55$ K) grows to, $\rho_\textrm{LCMO18}$(T $= 55$ K) $= 164$ K$\Omega$cm. In contrast, \NPMOsam, which has a much larger carrier concentration, has $\rho_\textrm{NPMO30}$(T $= 300$ K) $= 4 \Omega$cm, and $\rho_\textrm{NPMO30}$(T $= 55$ K) $= 1.15$ K$\Omega$cm. Both the systems however, despite such large differences in carrier concentration and resistivity, undergo the paramagnetic to ferromagnetic (Curie) transition at nearly the same temperature (\tc(LCMO18) $= 165$ K and \tc(NPMO30) $= 150$ K). Also, the temperatures below which the respective FMI states are realized (\tfmi), are nearly the same (\tfmi(LCMO18) $\approx 100$ K and \tfmi(NPMO30) $\approx 130$ K). To illustrate these aforementioned points, in figure~\ref{Fig1NPMO30LCMO18RhovsTComparison} we show the resistivity ($\rho$) as a function of temperature (T) of both systems for comparison; the respective paramagnetic to ferromagnetic (Curie) transition temperatures (\tc) are indicated by arrows. In this paper we report that despite having very different carrier concentration and resistivity as compared to \LCMOoneeight, the \NPMOsam system also shows collapse of magnetoresistance (MR) and a substantial electroresistance (ER) when the system enters the ferromagnetic insulating (FMI) state. This observation thus establishes that the phenomena is general, and may indeed be characteristic of the FMI state. 

We show that in the FMI region (T $\lesssim 75$ K) the MR essentially collapses to a very small value while the ER reaches its saturation value $\approx 100\%$. Furthermore, in this temperature region, the sample shows electric current induced switching of the resistance state where the resistivity can be switched from a high to a low value by switching the current from a low to a high value. The observation of a CER phenomena in absence of a CMR also distinguishes the FMI state from a conventional charge ordered insulating (COI) state which can be destabilized by both current and magnetic field~\cite{AsamitsuNATURE1997,GuhaPRB2000}.

We also note that, while the L $\equiv$ Pr and L $\equiv$ La based manganites have been investigated earlier from the point of view of electroresistance~\cite{ZhaoAPL2005,OdagawaPRB2004}, this is the first report of such phenomena for the Nd based manganite family. The experiments were carried out on a single crystal of \NPMOsam, that has a FMI phase below T $\approx 130$ K as determined from the resistivity vs temperature data.

\section{Experimental Details}
The single crystals of composition \NPMOsam (NPMO30) were grown by the flux-growth technique using PbO/PbF$_2$ solvent.  The melt batch was prepared from Nd$_2$O$_3$, MnCO$_3$, PbO and PbF$_2$. The crystals were grown in Pt crucible with charge to flux ratio of $1$:$6$ and PbO/PbF$_2$ ratio of $1$:$1.15$. The details of growth process have been given elsewhere~\cite{GhoshJMMM2003,GhoshJAP2004}. The crystals used have typical dimensions of $2\times2\times3$ mm$^3$. The crystals were characterized by single crystal X-Ray diffraction (space group P4/mmm) and the composition checked by energy dispersive X-Ray (EDAX) analysis and Inductively Coupled Plasma Atomic Emission Spectroscopy (ICPAES)~\cite{GhoshJMMM2003}. 

\begin{figure}[t]
\includegraphics{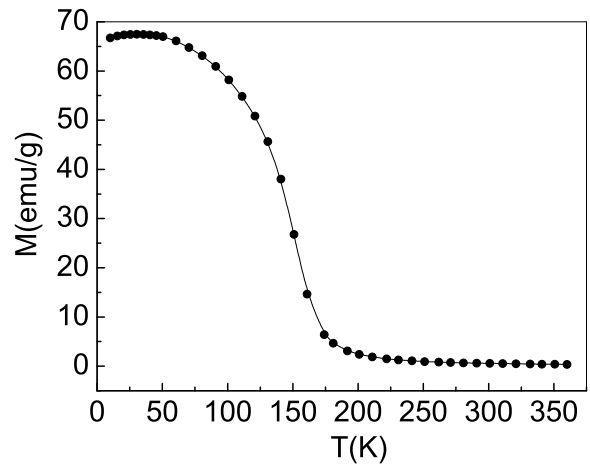}
\caption{\label{Fig2NPMO30MvsT}Magnetization (M) as a function of temperature (T) for \NPMOsam measured in a magnetic field, H $= 5$ T. The paramagnetic to ferromagnetic (Curie) transition at \tc $= 150$ K is evident.}
\end{figure}

The magnetization (M) as a function of temperature (T), measured by a SQUID magnetometer, is shown in figure~\ref{Fig2NPMO30MvsT}. The  sample has a paramagnetic to ferromagnetic (Curie) transition temperature, \tc $\approx 150$ K (as determined using a Curie-Weiss plot) and, a saturation moment of $3.8$ \mub per formula unit at T $= 20$ K in a magnetic field, H $= 5$ T.

Four linear gold contact pads were evaporated onto the crystal and Cu leads (of diameter $60$ $\mu$m) were  soldered onto the pads using an Ag-Sn alloy. The experiments were carried out by three methods: (1) by measuring resistance at fixed bias currents as a function of temperature, (2) by taking current-voltage (I-V) characteristics at different fixed temperatures, and (3) by measurement of the resistance in response to a two-level current pulse train. While all three methods yield the same data for a given set of parameters, the data reported here were taken mostly by the third method of pulsed current. In this technique, the levels of the high and low current values, the duration of each bias level as well as the number of pulses in the pulse train can be controlled. Most importantly, taking data in pulse mode reduces effect of Joule heating, if any, to a minimum. It was checked that there was no appreciable heating of the crystal during the measurements. Direct measurement of the rise of sample temperature, by attaching (thermally anchoring) a thermometer directly on top of the crystal, revealed that, for a steady high current bias (\ihigh $= 10$ mA), even at the lowest temperature, the rise of sample temperature is below $4$ K, and as the temperature is increased, it is negligible.

The experiments were carried out in a bath type cryostat. The pulsed bias resistivity measurements were performed by applying a two-level current pulse train to the sample and measuring the resulting voltage output in a standard 4-probe geometry.

The switching experiment was done by switching the measuring current between a high value and a low value and measuring simultaneously the voltage across the sample, from which the resistance was calculated. The choice of the particular low and high current values was based on measurements of the I-V characteristics. The low value was chosen to be one within the regime where the sample I-V characteristics are linear, and the high current value was chosen from within a regime where the I-V characteristics are strongly nonlinear. The switching time between high and low bias current values was approximately $100$ $\mu$s. In order to record the instantaneous response voltage profile of the sample upon switching the current bias level, the sampling time of each voltage measurement was set at a low value of $20$ ms; in addition, no filtering of the sampled data was performed. This enabled us to record correctly the transient response of the sample upon the application of current bias level. Again, at a particular bias level, the response time of the current source (the time required by the instrument to restore the bias to specified fixed value across a changing resistive load) was $\sim$ $30$ $\mu$s, which is $3$ orders of magnitude less than the time over which signal is sampled; therefore, from the perspective of measurement considerations, the transient response of the sample is indeed at constant bias (high or low). The (relative) time associated with each voltage measurement was determined via a synchronized oscillator with a precision better than $1$ ms.

\section{Results and Discussion}
\begin{figure}[t]
\includegraphics{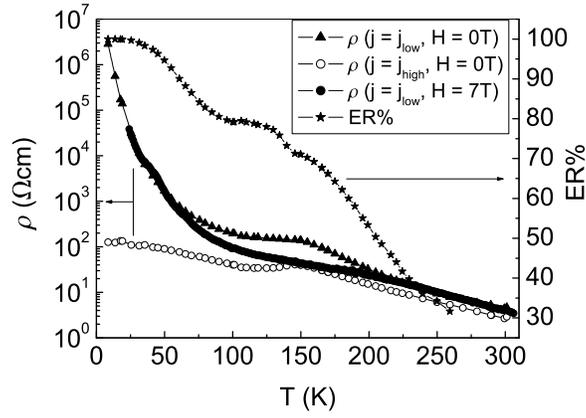}
\caption{\label{Fig3NPMO30I10uA10mA0T10uA7TRhoERvsT}The resistivity vs temperature ($\rho$-T) data obtained using low bias current density (\jlow) taken with and without a magnetic field (H $= 0$ T and $7$ T), and the $\rho$-T data taken with high current density (\jhigh) in H $= 0$ T. Electroresistance (ER) as a function of temperature is also plotted (\mbox{ER$\%$ = 100$\times$($\rho$(\jlow) - $\rho$(\jhigh))/$\rho$(\jlow)}). It may be noted that, for T $< 100$ K, the ER $\rightarrow$ 100$\%$, while the magnetoresistance, MR $\rightarrow$ 0$\%$.}
\end{figure}

In figure~\ref{Fig3NPMO30I10uA10mA0T10uA7TRhoERvsT} we show the resistivities ($\rho$) of \NPMOsam as a function of temperature (T) measured with two current densities (j) in absence of magnetic field (H $= 0$ T). The data were taken with current density, j = \jlow $\approx 6.7\times10^{-4}$ A/cm$^2$ (corresponding to a sample current, \ilow $= 10$ $\mu$A). At this value of j, the $\rho$ is independent of the measuring current density. The $\rho$ exhibits a peak at \tp $\approx 150$ K below which the sample enters a ferromagnetic metallic (FMM) phase. This FMM phase however, appears to be highly resistive and percolative in character. The behavior in this phase, of the $\rho$-T curve, suggests that the FMM phase may be coexisting with an insulating phase. The $\rho$ shows a small drop till T $\approx 130$ K, and below this temperature, with onset of the FMI phase, it increases rapidly following a mott variable range hopping relation~\cite{MottBOOK1993} for T $\lesssim 75$ K. In figure~\ref{Fig3NPMO30I10uA10mA0T10uA7TRhoERvsT} we also show the data taken using a high bias current density, \jhigh $\approx 6.7\times10^{-1}$ A/cm$^2$ (corresponding to a current, \ihigh $= 10$ mA). It is evident that there is a significant decrease in $\rho$ due to the (high) applied current. This current induced suppression of the resistivity is progressively enhanced as the temperature decreases, and at the lowest temperature measured, $\rho$(\jhigh)/$\rho$(\jlow) $< 5\times10^{-5}$.

In figure~\ref{Fig3NPMO30I10uA10mA0T10uA7TRhoERvsT} we also show the most important observation of the present investigation. Here, alongside the aforementioned low current zero magnetic field data, we show the $\rho$ measured using low current density (\jlow) in a magnetic field, H $= 7$ T. It is clear that there is substantial magnetoresistance (MR) in the temperature range $65$--$225$ K (attaining its highest value of $\approx 80\%$ at \tp $\approx 150$ K just as in other CMR single crystalline materials~\cite{RaoBOOK1998,TokuraBOOK2000TokuraRPP2006}. Furthermore as temperature decreases, the sample enters the FMI state at \tfmi $\approx 130$ K and the MR becomes very small ($\leq 20\%$) below $75$ K. The above observation is to be compared with the effect, on the sample, of a large bias current density: comparing this $\rho$-T data taken in a magnetic field, to the $\rho$-T data measured at high current density (\jhigh) makes evident that, while the magnetic field has a negligible effect on the resistivity, the current induces a substantial ($4$ orders of magnitude) depression of resistivity ($\rho$). In figure~\ref{Fig3NPMO30I10uA10mA0T10uA7TRhoERvsT} we also plot the electroresistance (ER) defined as,

\begin{equation}
\text{ER}\% \text{(T)}= 100\times \frac{\rho(\text{j}_\text{low}) - \rho(\text{j}_\text{high})}{\rho(\text{j}_\text{low})}\text{(T)}
\label{ER}
\end{equation}

as a function of temperature. It can be seen that, when the MR collapses in the FMI phase (T $\lesssim 75$ K) the ER picks up reaching a value of nearly 100$\%$ at the lowest temperature. Thus, this experiment demonstrates that there is an effective decoupling of the mechanism of ER and that of MR in the FMI phase. This result is rather distinct from previous results that used compositions in the ferromagnetic metallic (FMM)~\cite{WuPRL2001} phase or charge ordered insulating (COI)~\cite{AsamitsuNATURE1997,GuhaPRB2000} phase. Our observation therefore, clearly distinguishes the FMI state from the COI state. The former can be destabilized, to a more conducting state, only by a current while the latter can be destabilized by both current and magnetic field. 

\begin{figure}[t]
\includegraphics{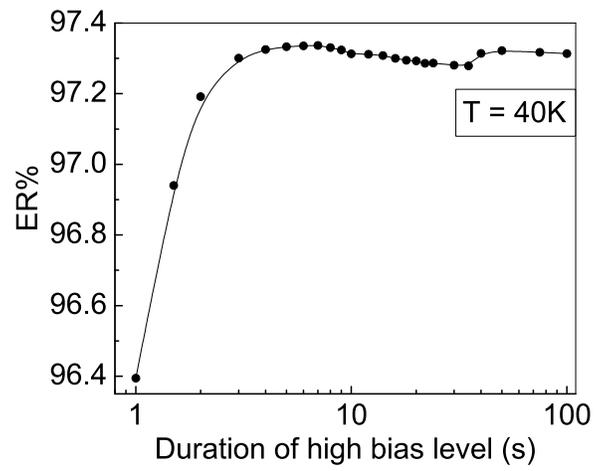}
\caption{\label{Fig4NPMO30ERvsPD}Absence of any substantial variation of electroresistance (ER) as the duration of high current bias level is changed in a typical pulsed current experiment.}
\end{figure}

To test the possibility of Joule heating being responsible for the above effect, we took data using the pulsed current technique, by varying the duty cycle. The results, shown in figure~\ref{Fig4NPMO30ERvsPD} are independent of the duty cycle. For instance, at T $= 40$ K, by varying the high current \lq\lq on\rq\rq time from $1$--$100$ s (i.e., over $2$ orders of magnitude), the measured resistivity changed by $< 0.5\%$, thus ruling out any significant sample heating, at least down to T $= 40$ K. Also, as has already been mentioned, the sample temperature rise (at the lowest measured temperature) due to passage of high bias current (\ihigh $= 10$ mA), as measured by directly attaching a thermometer on top of the sample was estimated to be $< 4$ K.

In figure~\ref{Fig5NPMO30IVInset10mA} we show the current vs voltage (I-V) characteristics taken on the sample at different temperatures (the data is rendered independent of particular sample dimensions by plotting it in specific units as current-density vs electric-field (j-E)). The nonlinear nature of the transport is clearly evident, and the data reveal a continuous decrease in $\rho$ ($=$ E/j) as j increases. The nonlinearity becomes progressively more severe with decreasing temperature. In fact, for T $< 25$ K, the j-E curve even shows a region of negative differential resistivity (NDR) where dE/dj $< 0$. We have earlier reported such an NDR effect in a charge ordered insulating (COI) sample of composition Pr$_{0.63}$Ca$_{0.37}$MnO$_3$~\cite{GuhaPRB2000,Guha1PRB2000}. Occurrence of such an NDR effect has also been reported  in COI state of Nd$_{0.5}$Ca$_{0.5}$MnO$_3$~\cite{GuhaAPL1999}. However, this is for the first time that NDR is reported in the ferromagnetic insulating (FMI) state. At present, however, there is no clear understanding of the NDR phenomena. 

Another notable feature of our data is the absence of hysteresis in the I-V (j-E) curves shown in figure~\ref{Fig5NPMO30IVInset10mA}. This is further evidence distinguishing the FMI state from the charge ordered (CO) state. This is so because, a CO state is characterized by a strong hysteresis in the j-E characteristics~\cite{GuhaPRB2000}. The hysteretic j-E curve in COI materials is often taken as a signature of electronic phase separation. It arises due to the kinetic effect, as determined by the potential barrier that separates the phases. Based on this observation of non-hysteretic j-E characteristics in the FMI state, we suggest that, whatever may be the underlying natures of the phases comprising the phase-separated state, the potential barrier between them, is low enough to have precluded the observation of any substantial kinetic effects in the time window of the present measurements. 

In figure~\ref{Fig6NPMO30T55KlnRhovslnj} we show representative data (at T $= 55$ K) of variation of resistivity ($\rho$) as a function of current density (j), plotted in a log-log (base $10$) scale. It is clear that there exists a threshold current density (\jth, indicated by an arrow) that separates two distinct $\rho$-j regimes, and that, in both regimes, the dependence of $\rho$ on j is a power law. Below the threshold current i.e., j $<$ \jth, $\rho$ is weakly dependent on j and in this region $\rho\propto$ j$^{\textrm{m}_1}$, with m$_1$ close to zero. However, for j $>$ \jth, a strong nonlinear regime sets in, and $\rho$ decreases strongly with increasing j.

The $\rho$-j data-sets then, taken at fixed temperatures, may be fitted to the relation 

\begin{equation}
\rho = \alpha_\text{n}\text{j}^{\text{m}_{\text{n}}} 
\label{powerlaw}
\end{equation}

where the subscripts n $= 1$ and $2$ denote the regimes j $<$ \jth and j $>$ \jth respectively. The thus obtained exponents m$_1$ and m$_2$, and the coefficients $\alpha_1$ and $\alpha_2$  are plotted in figure~\ref{Fig7NPMO30a1a2m1m2jthvsT} as a function of temperature (T). The systematic development of the nonlinear effects and the strong current dependence of the resistivity (which gives the CER) in the low temperature FMI phase can be seen. The CER is related to the strong nonlinear component that appears beyond the threshold current density \jth. It is evident from figure~\ref{Fig7NPMO30a1a2m1m2jthvsT}(a---d) that the nonlinear conductivity, as measured by $|$m$_2$ $|$ and $\alpha_2$ increases sharply at T $\approx 38$ K. In fact, m$_1$ and $\alpha_1$ also show a marked change at this temperature. In figure~\ref{Fig7NPMO30a1a2m1m2jthvsT}(e) we show the variation of the threshold current density (\jth) as a function of temperature (T). It can be seen that, while \jth has a shallow T dependence above T $= 38$ K (decreasing with decreasing temperature), there is a large jump, followed by a rapid drop below this temperature. In fact, all the parameters --- $\alpha_1$, $\alpha_2$, $m_1$, $m_2$ --- exhibit an anomaly at this temperature. It is suspected that this sharp anomaly may be caused by a phase transition type of phenomena setting in at T $\approx 38$ K. Detailed investigation of this is underway. The negative differential resistivity (NDR) seen in the j-E curves at low temperatures (see figure~\ref{Fig5NPMO30IVInset10mA}) typically appears after the system has undergone this transition. It appears that the CER is a manifestation of the charge current enhancing the transfer integral between the \eg electrons of the neighboring atoms, thereby enhancing the bandwidth and thus reducing the gap in the density of states (DoS), which in turn leads to a decrease of the resistivity~\cite{BergerJAP2001}.

In figure~\ref{Fig8NPMO30T130K42KRhojvst} we show representative current induced resistivity switching data. Figure~\ref{Fig8NPMO30T130K42KRhojvst}(a) shows the data at a temperature when the sample has just entered the ferromagnetic insulating (FMI) state (T $= 130$ K) and figure~\ref{Fig8NPMO30T130K42KRhojvst}(b) shows data when the sample is deep into the FMI state (T $= 42$ K). The switching of the resistance follows the switching of the current with a small drag of $\lesssim 2$--$3$ s, following which the full time independent value of ER (as plotted in figure~\ref{Fig3NPMO30I10uA10mA0T10uA7TRhoERvsT}) is realized. This rules out sample heating as that would have caused a change in the ER as the sample heats up.

The switching phenomena seen in the FMI samples is a strong manifestation of the CER effect. The switching data shows that the resistance of the sample follows the measuring current (almost) without any lag. This rules out any significant capacitive or charge storage effect, as that would have caused the voltage rise to lag behind the current.

\section{Conclusions}
In the present investigation we have established that it is possible to have colossal electroresistance (CER) in absence of a substantial magnetoresistance (MR) at least in one type of manganites, namely those that exhibit a low temperature ferromagnetic insulating (FMI) phase. We have observed this in two widely differing samples namely \NPMOsam (present investigation) and \LCMOoneeight~\cite{JainAPL2006}. This is an important and conclusive proof that the origin of the two phenomena of CER and CMR can be different. It is likely, though not necessary, that the FMI phase is electronically phase separated with ferromagnetic metallic (FMM) pockets in a matrix of insulators. The volume fraction of the insulating phase being in majority, results in the material being insulating. The passage of current destabilizes the insulating phase making it more metallic; this increase in the volume of the metallic phase reduces the resistivity. This scenario is distinct from the formation of ferromagnetic metallic filaments as has been seen in charge ordered insulating (COI) state~\cite{Guha1PRB2000,FiebigSCIENCE1998}. For FMI materials, a likely cause of the destabilization of the insulating phase can be a current induced enhancement of transfer integral between the electrons of neighboring \eg orbitals that form the conduction band. In recent years, a number of theories have been proposed where the spin alignment or transfer by a current have been envisaged~\cite{BergerPRB1996,SlonczewskiJMMM1996}. Though applicability of such ideas has not been explored in detail for manganite systems, we may propose a simple scenario based on such ideas. The spin alignment in the double-exchange (DE) scenario promotes conductivity by enhancing the hopping integral between \eg orbitals that constitute the conduction band. We propose that an enhanced current, in excess of what a simple equilibrium situation would allow, will also force spin alignment. This would in turn enhance the hopping integral, leading to increase of bandwidth and, to more states in the density of states (DoS) near the Fermi level, and thus to a suppression of resistivity.  This enhancement can enhance the bandwidth and reduce the gap in the DoS, leading therefore, to enhanced conductivity. In such a scenario, the CER and current induced switching can occur by an enhanced current density and it can indeed be de-linked from the CMR phenomena.

We also note the distinct nature of the FMI phase. This phase is interesting because, in a double-exchange system, ferromagnetism is always associated with a metal-like (d$\rho$/dT $> 0$) behavior. If the FMI phase is explained as a phase-separated state consisting of a FMM phase and an insulating phase with the FMM phase fraction below the percolation threshold, then one may still explain the FMI phase within the framework of the double-exchange mechanism. However, a nearly full saturation magnetization of $3.8$ \mub per formula unit is difficult to explain if the FMM phase is in minority. The quandary arises from the nature of the insulating phase. The absence of any substantial MR suggests that the insulating nature of the FMI phase is distinct from that of COI phases~\cite{GuhaPRB2000}.

In summary, the present investigation has established two important and distinctive features of ferromagnetic insulating (FMI) phase of manganites. First, that the FMI phase shows no MR although it shows a CER. The CER is associated with strong nonlinear conductivity effects like current induced resistivity switching as well as negative differential resistivity. This observation points to different origins for the two phenomena of CER and CMR. Second, the FMI phase, as an insulating state is distinct from the charge ordered insulating (COI) state which can be destabilized by both a magnetic field as well as by an electric field, unlike the FMI phase.

\clearpage

\begin{acknowledgments}
One of the authors (H.J.) thanks the Council of Scientific and Industrial Research (CSIR), India, for a fellowship. Another author (A.K.R.) thanks the Department of Science and Technology (DST), India, for a sponsored project. 
\end{acknowledgments}

\clearpage

\section*{List of Figure Captions}
\begin{figure}[t]
\includegraphics{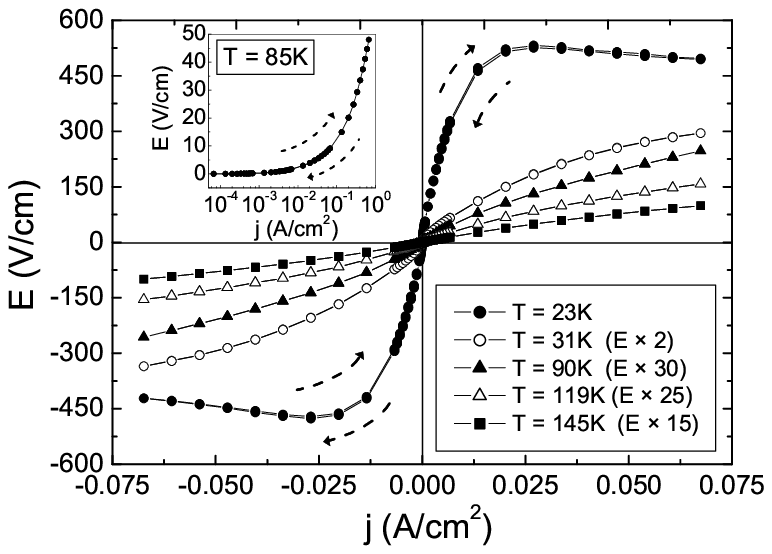}
\caption{\label{Fig5NPMO30IVInset10mA}Current vs voltage (I-V) characteristics of \NPMOsam at few representative temperatures (T) as indicated (the data is plotted in specific units as current-density vs electric-field (j-E)). A representative semi-log plot in inset demonstrates the absence of hysteresis in j-E characteristics over $4$ orders of magnitude variation in j. The T $= 23$ K j-E characteristic, exibiting the negative differential resistivity (NDR) feature is noteworthy. (Note the scaling of E by the numerical factors mentioned in brackets.)}
\end{figure}

\begin{figure}[t]
\includegraphics{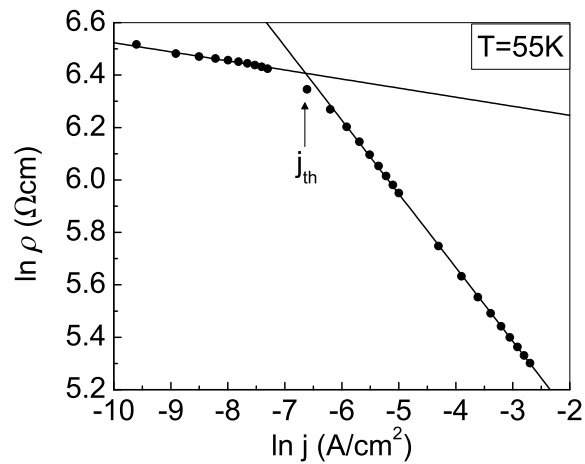}
\caption{\label{Fig6NPMO30T55KlnRhovslnj}A representative log-log plot showing two regimes of power law variation of resistivity ($\rho$) as a function of bias current density (j) at temperature, T $= 55$ K. The threshold current density (\jth) is indicated. The lines are least squares best fits in the two regimes. Such analysis was used to obtain the parameters --- $\alpha_1$, $\alpha_2$, m$_1$, m$_2$ --- as are plotted in figure~\ref{Fig7NPMO30a1a2m1m2jthvsT}.}
\end{figure}

\begin{figure}[t]
\includegraphics{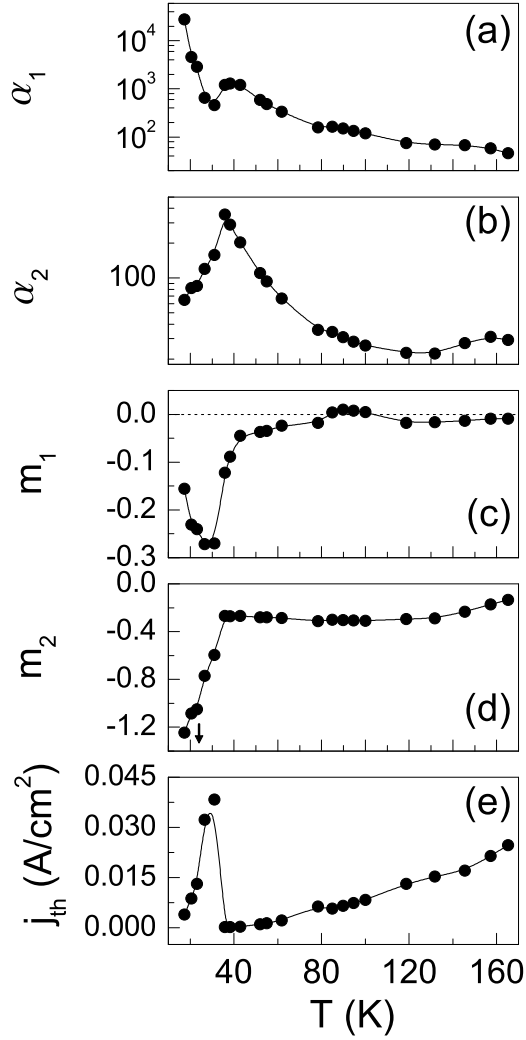}
\caption{\label{Fig7NPMO30a1a2m1m2jthvsT}Variation of (a) $\alpha_1$, (b) $\alpha_2$, (c) m$_1$, (d) m$_2$, and  (e) \jth, as a function of temperature (T) in the ferromagnetic insulating state. One may notice the rapid decrease in m$_2$ for T $\lesssim 38$ K. The negative differential resistivity (NDR) regime (m$_2 < -1$) is realized for T $\lesssim 25$ K (indicated by arrow in (d)). See also equation~\ref{powerlaw} and figure~\ref{Fig5NPMO30IVInset10mA}.}
\end{figure}

\begin{figure}[t]
\includegraphics{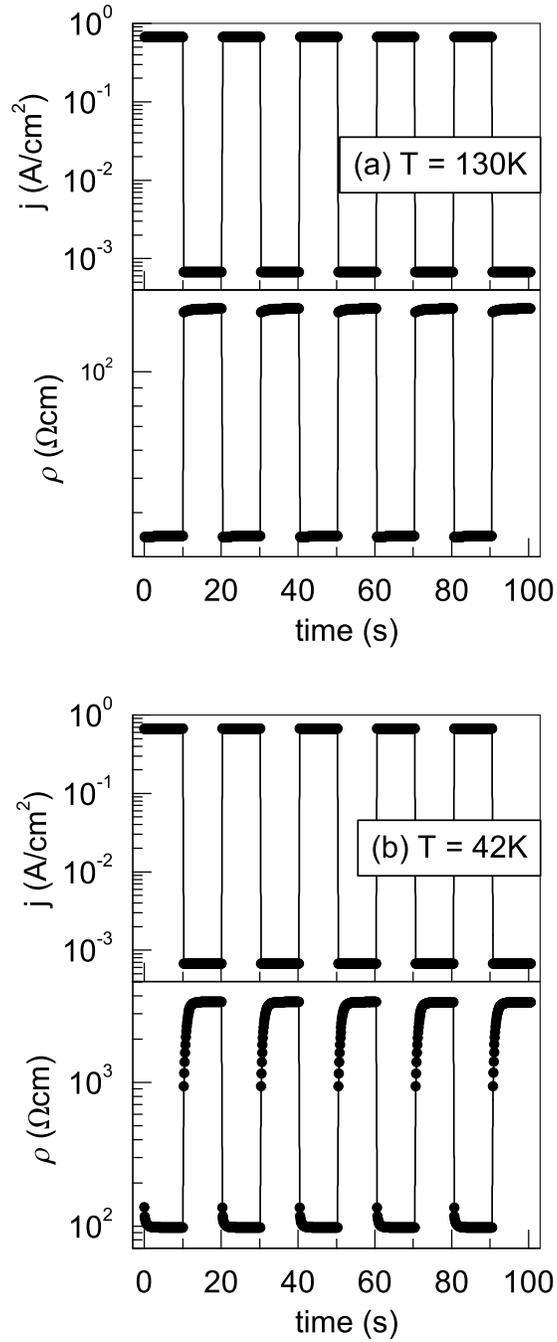}
\caption{\label{Fig8NPMO30T130K42KRhojvst}Representative data-sets showing switching of resistivity ($\rho$) upon application of a current density (j) pulse train at, (a) T $= 130$ K (just within the ferromagnetic insulating (FMI) state), and (b) T $= 42$ K (deep in the FMI state).}
\end{figure}

\begin{itemize}
\item[FIG. 1]{Comparison of resistivities ($\rho$) of \NPMOsam and \LCMOoneeight as a function of temperature (T). The paramagnetic to ferromagnetic (Curie) transition temperatures (\tc), indicated by arrows, are $150$ K and $165$ K respectively.}
\item[FIG. 2]{Magnetization (M) as a function of temperature (T) for \NPMOsam measured in a magnetic field, H $= 5$ T. The paramagnetic to ferromagnetic (Curie) transition at \tc $= 150$ K is evident.}
\item[FIG. 3]{The resistivity vs temperature ($\rho$-T) data obtained using low bias current density (\jlow) taken with and without a magnetic field (H $= 0$ T and $7$ T), and the $\rho$-T data taken with high current density (\jhigh) in H $= 0$ T. Electroresistance (ER) as a function of temperature is also plotted (\mbox{ER$\%$ = 100$\times$($\rho$(\jlow) - $\rho$(\jhigh))/$\rho$(\jlow)}). It may be noted that, for T $< 100$ K, the ER $\rightarrow$ 100$\%$, while the magnetoresistance, MR $\rightarrow$ 0$\%$.}
\item[FIG. 4]{Absence of any substantial variation of electroresistance (ER) as the duration of high current bias level is changed in a typical pulsed current experiment.}
\item[FIG. 5]{Current vs voltage (I-V) characteristics of \NPMOsam at few representative temperatures (T) as indicated (the data is plotted in specific units as current-density vs electric-field (j-E)). A representative semi-log plot in inset demonstrates the absence of hysteresis in j-E characteristics over $4$ orders of magnitude variation in j. The T $= 23$ K j-E characteristic, exibiting the negative differential resistivity (NDR) feature is noteworthy. (Note the scaling of E by the numerical factors mentioned in brackets.)}
\item[FIG. 6]{A representative log-log plot showing two regimes of power law variation of resistivity ($\rho$) as a function of bias current density (j) at temperature, T $= 55$ K. The threshold current density (\jth) is indicated. The lines are least squares best fits in the two regimes. Such analysis was used to obtain the parameters --- $\alpha_1$, $\alpha_2$, m$_1$, m$_2$ --- as are plotted in figure~\ref{Fig7NPMO30a1a2m1m2jthvsT}.}
\item[FIG. 7]{Variation of (a) $\alpha_1$, (b) $\alpha_2$, (c) m$_1$, (d) m$_2$, and  (e) \jth, as a function of temperature (T) in the ferromagnetic insulating state. One may notice the rapid decrease in m$_2$ for T $\lesssim 38$ K. The negative differential resistivity (NDR) regime (m$_2 < -1$) is realized for T $\lesssim 25$ K (indicated by arrow in (d)). See also equation~\ref{powerlaw} and figure~\ref{Fig5NPMO30IVInset10mA}.}
\item[FIG. 8]{Representative data-sets showing switching of resistivity ($\rho$) upon application of a current density (j) pulse train at, (a) T $= 130$ K (just within the ferromagnetic insulating (FMI) state), and (b) T $= 42$ K (deep in the FMI state).}
\end{itemize}

\clearpage

\clearpage

\clearpage

\clearpage

\clearpage

\clearpage

\clearpage

\clearpage

\end{document}